\documentstyle[prl,aps]{revtex}
\begin{document}

\twocolumn[\hsize\textwidth\columnwidth\hsize
            \csname @twocolumnfalse\endcsname

\narrowtext
]
{\bf Comment on ``Off-stoichiometry mechanism of the isotope effect
in manganites''}
~\\
~\\
In a recent paper, Nagaev \cite{Nagaev} cited the unpublished paper by
Franck {\em et al.} to support his theoretical model for the mechanism
of the giant isotope effect observed in
La$_{1-x}$Ca$_{x}$MnO$_{3+y}$ ($x$ = 0.20, $y$ $>$ 0) \cite{ZhaoNature}.
His model suggests that the off-stoichiometric oxygen content depends
strongly on the oxygen isotope mass, which leads to a giant
oxygen-isotope effect.
Here I show that his theoretical model is not consistent with any
experimental results (even the results recently published by Franck {\em et
al.} \cite{Franck}), and his estimate of polaronic
bandwidth is wrong due to his misuse of
polaronic theories.

Nagaev cited the unpublished data by Franck {\em et al.} to support
his claim.  As a matter of fact, Franck {\em et al.} have discarded the
data Nagaev cited, and only published the reliable data which clearly
show that the oxygen contents of both isotope samples are the same
even if the samples are off-stoichiometric and have a giant
oxygen-isotope shift \cite{Franck}.
Actually  the giant oxygen-isotope effect
for the off-stoichiometric samples ($x$ = 0.20, $y$ $>$ 0) is due to
the fact that the
samples are metallic and close to the boundary of the metal-insulator
transition.
The reduced isotope effect in the stoichiometric samples ($x$ = 0.20,
$y$ = 0) \cite{Franck} may be tied with the fact that the samples are
ferromagnetic
insulators at low temperatures \cite{Hundley,Okuda}. The reduced
pressure and isotope effects observed in the insulating ferromagnetic
phase \cite{ZhaoPRB} should be associated with the reduction of the
ferromagnetic coupling
contributed from charge carriers.

There are other experimental facts which can further argue against the model
by Nagaev \cite{Nagaev}. According to his argument \cite{Nagaev}, the
isotope effect in more nonstoichoimetric samples should be larger. This is
in contradiction
with experiment. The stoichoimetric
(La$_{0.25}$Pr$_{0.75}$)$_{0.7}$Ca$_{0.3}$MnO$_{3}$ shows a
very large isotope effect \cite{Balagurov}, while the very
nonstoichoimetric (LaMn)$_{0.945}$O$_{3}$ has a rather small
isotope effect \cite{ZhaoPRL}.

If one carefully examines the theoretical calculation in Ref.\cite{Nagaev},
one will find that the excess oxygen content for the $^{16}$O sample
should be lower than for
the $^{18}$O sample by about 4$\%$ at temperatures $T >
\omega$, where $\omega$ is the highest phonon frequency, which is 950
K in LaMnO$_{3}$ \cite{Abr}.  Because the oxygen-isotope exchange was
carried out
at $T >$ 1200 K ($>$ $\omega$), and because the oxygen content in
La$_{0.8}$Ca$_{0.2}$MnO$_{3+y}$ remains unchanged below 1000
K ($>$ $\omega$) according to our thermogravimetric analysis, the
calculation of Ref.~\cite{Nagaev} actually suggests that the excess
oxygen content $y$ for the $^{16}$O sample should be lower than for
the $^{18}$O sample by about 4$\%$.  With $y$ = 0.02 
(Ref.~\cite{Hundley}), one finds that the
oxygen content for the $^{16}$O sample is lower than for the $^{18}$O sample
by 0.0008. Thus, the predicted isotope effect is negligibly small and
has an opposite sign as what observed.

His argument against strong electron-phonon coupling in
manganites is lack of scientific grounds. Millis {\em et al.}
\cite{Millis} clearly
showed that the double-exchange alone cannot explain the resistivity
peak at $T_{C}$ and thus proposed that strong electron-phonon coupling
should be involved to describe the basic physics of the material. The
experimental evidence for the strong electron-phonon coupling in
manganites is overwelming \cite{MillisNature}.

The estimate of the polaronic
bandwidth in Ref.~\cite{Nagaev} is incorrect. Nagaev misuses the 
small polaron theory, applying
the Holstein $nonadiabatic$ expression for the polaron bandwidth to
the $adiabatic$ region of the parameters: the bare bandwidth $W_0$= 1eV and
the characteristic phonon frequency $\omega = 0.01$ eV. It is well 
known \cite{Alex94} that the nonadibatic
expression overestimates the actual bandwidth by many orders of
magnitude, if it is applied to a wrong region of parameters. Actually
the $\omega = 0.01$eV
used by Nagaev in his estimate of the polaron bandwidth is inconsistent with
the frequency ($\omega
\simeq 0.1$ eV) used to justify his theory of the isotope
effect in the same paper \cite{Nagaev}. With  $\omega \simeq 0.1$ eV, 
and the  polaron
binding energy $E_p=0.5$ eV, the polaron bandwidth is not
reduced so dramatically, as Nagaev suggests, but only about one
order of magnitude, when the right expression is applied
\cite{Alex94}.

Finally, Nagaev's claim that small polarons are inevitably localized
in the random potential is false. All states of the polaronic
band might be localized if the fluctuation energy, $F$ is about 5
times larger than the polaron half-bandwidth, $D$ \cite{textbook}. 
Even with the
impurity density as high as $n_{im} = 0.3$ per cell, one obtains $F/D
\simeq 0.5$ using the polaron mass, $m^{*}=10~m_e$. Hence, $F/D$ is about one
order of magnitude below the critical ratio for
localization of all polaronic states.
~\\
~\\
Guo-meng Zhao \\
Department of Physics and Texas Center for Superconductivity,
University of Houston, Houston, Texas 77204.\\

\end{document}